\documentclass[12pt,a4paper,reqno]{article}
\pdfoutput=1
\usepackage{amsmath}
\usepackage{amssymb}
\usepackage{bbm}
\usepackage{graphicx}
\usepackage{hyperref}

\hypersetup{colorlinks,linkcolor=blue,citecolor=red,urlcolor=blue,%
bookmarksopen,pdftitle={Free Lunch from T-Duality},%
pdfauthor={Ulrich Theis (FSU Jena)}}

\allowdisplaybreaks

\renewcommand{\t}{\theta}
\renewcommand{\#}{\mskip-1mu}
\newcommand{\+}{\mskip2mu} 
\newcommand{\e}{\mathrm{e}}
\newcommand{\I}{\mathrm{i}}
\newcommand{\cL}{\mathcal{L}}
\newcommand{\fR}{\mathbb{R}}
\newcommand{\al}{\alpha}
\newcommand{\be}{\beta}
\newcommand{\ga}{\gamma}
\newcommand{\de}{\delta}
\newcommand{\vphi}{\varphi}
\newcommand{\vt}{\vartheta}
\newcommand{\V}{\xi}
\newcommand{\ep}{\varepsilon}
\newcommand{\eps}{\epsilon}
\newcommand{\si}{\sigma}
\newcommand{\p}{\partial}

\newcommand{\half}{\tfrac{1}{2}}
\newcommand{\diag}{\text{diag}\mskip1mu}
\newcommand{\lp}{\begin{pmatrix}}
\newcommand{\rp}{\end{pmatrix}}
\newcommand{\frc}[2]{\frac{\raisebox{-2pt}{$#1$}}{#2}}
\newcommand{\arxiv}[1]{\href{http://arxiv.org/abs/#1}%
{\texttt{arXiv:#1}}}

\begin{document} 

\begin{center}
 {\large\bfseries Free Lunch from T-Duality} \\[3ex]
 {\large Ulrich Theis} \\[1ex]
 \slshape Institute for Theoretical Physics,
  Friedrich-Schiller-University Jena, \\ Max-Wien-Platz 1,
  D--07743 Jena, Germany \\
 \href{mailto:Ulrich.Theis@uni-jena.de}{Ulrich.Theis@uni-jena.de}
\end{center}

\vspace{5ex} \hrule \vspace{3ex}

We consider a simple method of generating solutions to Einstein gravity coupled to a dilaton and a 2-form gauge potential in $n$ dimensions, starting from an arbitrary $(n-m)$-dimensional Ricci-flat metric with $m$ commuting Killing vectors. It essentially consists of a particular combination of coordinate transformations and T-duality and is related to the so-called null Melvin twists and TsT transformations. Examples obtained in this way include two charged black strings in five dimensions and a finite action configuration in three dimensions derived from empty flat space. The latter leads us to amend the effective action by a specific boundary term required for it to admit solutions with positive action. An extension of our method involving an S-duality transformation that is applicable to four-dimensional seed metrics  produces further nontrivial solutions in five dimensions.

\vspace{3ex} \hrule \vspace{5ex}

\section{Introduction} 

One of the most attractive features of string theory is that it gives rise to gravity: the low-energy effective action for the background fields on a string world-sheet is diffeomorphism invariant and includes the Einstein-Hilbert term of general relativity. This effective action is obtained by integrating renormalization group beta functions whose vanishing is required for Weyl invariance of the world-sheet sigma model and can be interpreted as a set of equations of motion for the background fields. Solutions to these equations determine spacetimes in which the string propagates.

In the case of the closed bosonic string the massless background fields consist of a metric, a two-form (antisymmetric tensor) gauge potential and a scalar field, the dilaton. Neglecting the central charge term that only vanishes in 26 dimensions, their low-energy effective action is interesting in its own right without connection to string theory as a particular theory of matter coupled to gravity. In five dimensions especially the tensor can be dualized into a more conventional vector field, such that one obtains Einstein-Maxwell-dilaton theory. It is with this point of view in mind that we offer two computationally very simple methods of constructing nontrivial solutions to the equations of motion of this theory. It makes use of the target space duality symmetry (T-duality for short) of the corresponding world-sheet sigma model under inversion $R\rightarrow 1/R$ of the radius of a compact dimension (for a review see e.g.\ \cite{GPR}).

The action of T-duality on the background fields of the bosonic string was worked out to one-loop order in \cite{B1,B2}. As we briefly review in the next section, these Buscher rules map one solution of the equations of motion to another. Since they mix components of the metric and matter fields, the resulting spacetime geometry depends on the choice of coordinates used to formulate the seed solution. Suitable coordinate transformations/field redefinitions can be employed in combination with T-duality to derive new solutions entirely different from the original ones (though equivalent from the world-sheet point of view).

Consider for instance\footnote{I learned of this example from Martin Ro\v{c}ek \cite{RV}.} the flat metric on $\fR^2$ and pass from Cartesian to polar coordinates: $ds^2=dr^2+r^2 d\vphi^2$. Upon dualization of $\vphi$ into an angle $\t$ the Buscher rules yield the metric $ds^2=dr^2+r^{-2} d\t^2$, which is curved with curvature scalar $R=-4/r^2$. The field equations are solved thanks to a nontrivial dilaton $\Phi=-\ln r$. The methods we present in this paper expand on this observation by essentially performing coordinate transformations that produce off-diagonal components of the metric which after dualization give rise to a nontrivial tensor field in addition to the dilaton. More concretely, we use as input an arbitrary $(n-m)$-dimensional Ricci-flat metric that admits $m$ commuting Killing vectors, which obviously solves the vacuum field equations. We then extend spacetime by $m$ dimensions on which we perform T-duality transformations after a field redefinition that mixes the metric components along the old and new directions (i.e., couples the world-sheet scalar fields). The role of the isometries is to ensure the applicability of T-duality after the field redefinition. The result is a metric with (at least) $2m$ isometries and nontrivial matter fields living in $n$ dimensions.

When regarded as a procedure performed on a two-dimensional sigma model, our approach fits into a more general framework of certain $p$-form gauge theories, which gave us the idea to apply it to gravity in the first place: In \cite{HK} Henneaux and Knaepen generalized so-called Freedman-Townsend models \cite{FT} of self-interacting $(D-2)$-forms $\t_\al$ in $D$ dimensions to include couplings to further form fields $X^a$. We observed in \cite{T} that upon dualization of the $\t_\al$ into scalar fields $\phi^\al$ the latter can be decoupled from the $X^a$ under certain conditions by means of a local field redefinition, even though no such redefinition exists that could decouple the $\t_\al$ from the $X^a$ in the dual formulation on the level of the action.\footnote{All scattering amplitudes vanish, however, which resolves the apparent paradox that had us confused in \cite{T}.} In $D=2$ all forms are reduced to scalars and we obtain a duality between two sigma models, with the $(X^a,\t_\al)$ model having a curved target space even if the dual $(X^a,\phi^\al)$ model is flat. The $0$-forms $\t_\al$, regarded as world-sheet scalar fields, parametrize the $m$ additional dimensions mentioned above. The condition under which the decoupling of the dual scalars $\phi^\al$ occurs translates into isometries of the metric that provides the kinetic function of the $n-m$ $0$-forms $X^a$.

Of course, generating further solutions by T-dualizing existing ones is a straightforward procedure that would hardly warrant a publication. It becomes interesting, however, if additional steps such as ours produce highly nontrivial solutions out of seed solutions that are as simple as possible --- ideally trivial as in the example above, which then amounts to solving the Einstein equations without ever having to solve a single differential equation. Below we obtain several such nontrivial examples from empty flat space. Thus, the proverbial free lunch does exist after all.

Previous examples of deriving nontrivial solutions from (Ricci-) flat space by a combination of duality and coordinate transformations can be found for instance in \cite{HHS,KT,HT,AG}. In fact, when restricted to the simplest case of the flat metric in Cartesian coordinates with one isometry generated by an antisymmetric matrix as input and one corresponding additional dimension, our solutions reduce to ones that have been found before in \cite{AG} by means of what is now called a null Melvin twist in the literature \cite{GHHLR}. The case of nontrivial metrics extended by one dimension is a subset of solutions derived in \cite{ABDR} by the same null Melvin twist.\footnote{I am grateful to Oliver DeWolfe for pointing this out to me.} The first three examples below fall into this class, whereas the following ones are more general and cannot be obtained by a null Melvin twist.

In particular, the last two examples are derived by means of an extension of the method described above that can be applied to Ricci-flat metrics in four dimensions with an isometry and produces five-dimensional static spacetimes (if the seed metric has Euclidean signature) with nontrivial matter. It makes a detour through six dimensions, where the dual of a 2-form gauge potential is again a 2-form, and includes an S-duality transformation of the dilaton and a Weyl rescaling of the metric. After dualizing the tensor into a vector field, we obtain in this way novel solutions to Einstein-Maxwell-dilaton gravity in five dimensions.

We should remark that we do not regard our methods as a means to generate consistent bosonic string backgrounds. For one, we only solve the string equations to lowest order in the $\alpha'$-expansion, and it is unlikely that any of these field configurations are exact to all orders. Another reason is that we are neglecting the central charge term in the dilaton anomaly coefficient since we are very much interested in solutions in dimensions less than 26, which in our approach would require a nontrivial seed solution with nonvanishing matter. Nevertheless, we shall occasionally refer to the solutions as `backgrounds'.

In the next section we set up the theory and recall how T-duality maps solutions to solutions. We then point out an unsatisfactory property of the Euclidean $O(\alpha'{}^0)$ effective action of the bosonic string (without central charge term) as it is usually found in the literature: it is \emph{negative} semi-definite. As we explain in section \ref{sec:BdTerm}, this problem can be cured by including a suitable boundary term in the action. Such a modification is admissible as it leaves the beta function equations unchanged. The particular term we add is also natural in that it compensates for a total derivative that arises in passing from string to Einstein frame by means of a Weyl rescaling of the metric. This term is usually dropped, but can be nonvanishing and then destroys the positivity of the Euclidean matter action. In our first example, a three-dimensional geometry with finite action, we find that the boundary term has the effect of flipping the sign of the latter.

In section \ref{sec:Sol} we explain our main method of generating solutions, which we present in two equivalent versions formulated in different sets of coordinates. In the following section \ref{sec:Ex} we then apply our method to five seed solutions. The first two consist of nothing more than the flat metric on $\fR^2$ in Cartesian coordinates, once with Euclidean signature and once with Lorentzian one. The former leads to the aforementioned finite action solution, while the latter results in a three-dimensional spacetime with timelike singularities somewhat similar to those appearing in the Reissner-Nordstr\"om black hole. The third example is obtained from the Schwarzschild metric as input and describes a magnetically charged black string in five dimensions previously found in \cite{HHS} by other means. We also construct a corresponding electrically charged black string. The fourth example is a two-parameter deformation of six-dimensional empty flat space. From this we derive our second method of generating solutions in five dimensions, for which we provide two more examples: one again starting from four-dimensional flat space and one using the Euclidean Taub-NUT metric as input. The latter produces a static spacetime permeated by magnetc charge whose spatial slices are deformed Taub-NUT geometries. Finally, we conclude in section~\ref{sec:Con}.

\section{The Model and T-Duality} 
\label{sec:Tduality}

Consider the following $n$-dimensional action for a metric $G_{IJ}(X)$, antisymmetric tensor $B_{IJ}(X)$ and dilaton $\Phi(X)$ in the string frame:\footnote{We follow the conventions of Polchinski \cite{P} with $2\pi\alpha'=1$ and $2\kappa_0^2=1$.}
 \begin{equation} \label{S}
  S_n = \pm \int\! d^{\+n}\!X\, \sqrt{G\,}\+ \e^{-2\Phi} \Big( R + 4\, \p_I \Phi\, \p^I \Phi - \frc{1}{12} H_{IJK} H^{IJK} \Big) + S^b_n\ ,
 \end{equation}
where $H_{IJK}=3\+\p_{[I}B_{JK]}$, and the overall sign depends on the signature of the metric (plus in the Lorentzian case). Here, $S_n^b$ is a boundary term that we specify below. The linear combinations \cite{CFMP}
 \begin{align}
  2\pi \bar{\beta}^G_{IJ} & = R_{IJ} + 2\+ \nabla_{\!I} \p_J \Phi - \frc{1}{4} H_{IKL} H_J{}^{KL} = 0 \label{betaG} \\[2pt]
  2\pi \bar{\beta}^B_{IJ} & = - \frc{1}{2}\, \e^{2\Phi}\, \nabla^K \big( \e^{-2\Phi} H_{IJK} \big) = 0 \label{betaB} \\[2pt]
  2\pi \bar{\beta}^\Phi{}' & = - \frc{1}{2} \nabla^I \p_I \Phi + \p_I \Phi\, \p^I \Phi - \frc{1}{24} H_{IJK} H^{IJK} = 0 \label{betaP}
 \end{align}
of the equations of motion of $S_n$ correspond, up to a constant $n$-dependent central charge term missing in $\bar{\beta}^\Phi{}'$, to the vanishing of the Weyl anomaly coefficients (related to the one- and two-loop beta functions) of a nonlinear sigma model on a two-dimensional curved world-sheet with action
 \begin{equation}
  L_\si = \frc{1}{2}\, \Big( G_{IJ}\# *\! dX^I\! \wedge dX^J - B_{IJ}\, dX^I\! \wedge dX^J - \frc{1}{2\pi}\, \Phi\# *\!\# R^{(2)} \Big)\ .
 \end{equation}

If coordinates $X^I=(X^a,\t)$ with $a=1,\dots,n-1$ can be chosen such that the background fields in $L_\si$ do not depend on $\t$, then one may dualize $\t$ into another scalar $\phi$. To leading order the background fields transform under this duality according to the Buscher rules \cite{B1,B2} as
 \begin{gather}
  \tilde{G}_{ab} = G_{ab} - G_{\t\t}^{-1} \big( G_{a\t} G_{\t b} + B_{a\t} B_{\t b} \big)\ ,\quad \tilde{G}_{a\phi} = G_{\t\t}^{-1} B_{a\t}\ ,\quad \tilde{G}_{\phi\phi} = G_{\t\t}^{-1} \notag \\[4pt]
  \tilde{B}_{ab} = B_{ab} - G_{\t\t}^{-1} \big( G_{a\t} B_{\t b} + B_{a\t} G_{\t b} \big)\ ,\quad \tilde{B}_{a\phi} = G_{\t\t}^{-1}\, G_{a\t} \notag \\[4pt]
  \tilde{\Phi} = \Phi - \half \ln G_{\t\t}\ . \label{Buscher}
 \end{gather}
The anomaly coefficients \eqref{betaG}--\eqref{betaP} transform homogeneously under T-duality. This essentially follows from the renormalization group flow of the dual background fields obtained by regarding them as functions of a renormalization scale $\mu$ and applying the logarithmic derivative $\mu\,d/d\mu$ to the relations \eqref{Buscher}. Using the definition of the beta functions $\beta^G_{IJ}=\mu\,dG_{IJ}/d\mu$ etc.\ one obtains in this way their duality transformations \cite{H}. The same hold for the anomaly coefficients. We conclude that, by virtue of the homogeneity, T-duality maps solutions to the equations of motion of $S_n$ to solutions. Note that this is the case whether or not the coordinate $\t$ that we dualize parametrizes a compact dimension. In the following we will dualize several scalars $\t_\al$ at once, for which we explicitly determine the transformed background metric and tensor in section \ref{sec:Sol}, while the transformation of the dilaton we just quote from the literature.

\section{The Boundary Term} 
\label{sec:BdTerm}

Let us now explain the inclusion of the boundary term in the string frame action \eqref{S}. It is given by
 \begin{equation} \label{Sb}
  S^b_n = \mp\+ 2\, \frc{n-1}{n-2}\, \int\! d^{\+n}\!X\, \sqrt{G\,}\, \nabla^I \p_I\, \e^{-2\Phi}
 \end{equation}
and can be nonzero for metrics and dilatons that do not fall off sufficiently fast at infinity. It is needed for the theory to admit solutions with positive Euclidean action. To see this, note that using the equations of motion \eqref{betaG} and \eqref{betaP} we may write the terms in brackets in \eqref{S} as
 \begin{equation}
  R + 4\, \p_I \Phi\, \p^I \Phi - \frc{1}{12} H_{IJK} H^{IJK} = \frc{1} {3} H_{IJK} H^{IJK}\ ,
 \end{equation}
which implies that the usual Euclidean action without boundary term is negative semi-definite for \emph{any} solution. $S^b_n$ is  chosen such as to cancel the total derivative arising in the Weyl rescaling
 \begin{equation} \label{Weyl}
  \hat{G}_{IJ} = \e^{-4 \Phi / (n-2)}\, G_{IJ}
 \end{equation}
of the curvature scalar:
 \begin{equation}
  \sqrt{G\,}\+ \e^{-2\Phi} R = \sqrt{\#\hat{G}\,} \Big( \hat{R} - 4\, \frc{n-1}{n-2}\, \p_I \Phi\, \hat{\p}^I \Phi - 4\, \frc{n-1}{n-2}\, \hat{\nabla}^I \p_I \Phi \Big)\ .
 \end{equation}
The result is an Einstein frame action
 \begin{equation} \label{S_E}
  S_n = \pm \int\! d^{\+n}\!X\, \sqrt{\#\hat{G}\,} \Big( \hat{R} - \frc{4}{n-2}\, \p_I \Phi\, \hat{\p}^I \Phi - \frc{1}{12}\, \e^{-8 \Phi / (n-2)}\+ H_{IJK} \hat{H}^{IJK} \Big)
 \end{equation}
with manifestly positive semi-definite matter terms in the case of Euclidean signature\footnote{Provided the fields are real. Dualization in Euclidean spaces can result in imaginary fields with negative kinetic terms, as in our last examples.} and \emph{no} boundary term.

Of course, this is no guarantee that the action is always nonnegative, since the Einstein-Hilbert term is unbounded and may dominate over the matter terms. But positive action solutions now exist at least in special cases, as both the first example below and the following argument \cite{TV} show: In four dimensions one can derive a Bogomol'nyi bound $L_4\geq-\hat{R}+4\,\p_I\Phi\,\hat{\p}^I\Phi$ on the Euclidean action in Einstein frame that is saturated by fields satisfying $\hat{\p}^I\e^{\+2\Phi}=-\tfrac{1}{6}\,\ep^{IJKL}H_{JKL}$. As these have vanishing energy-momentum tensor, the metric must be Ricci-flat and we obtain $L_4\geq 0$. This reasoning would not apply if we didn't compensate for the total derivative in the Weyl-rescaled curvature scalar.

\section{The Solutions} 
\label{sec:Sol}

Let us first recall the generalized Freedman-Townsend models of \cite{HK} (extended by a topological term introduced in \cite{BT} that naturally occurs in supersymmetric versions). They describe interactions of $(D-2)$-forms $\t_\al$ and $p$-forms $X^a$ in $D$ dimensions, which are conveniently written in first-order form using auxiliary 1-forms $V^\al$:
 \begin{equation} \label{HKmodel}
  L_1 = \frc{1}{2}\, \big( \de_{\al\be} *\!V^\al\! \wedge V^\be + 2\, \t_\al \wedge G^\al + g_{ab} *\! F^a\! \wedge F^b + b_{ab}\, F^a\! \wedge F^b \big)
 \end{equation}
with field strengths
 \begin{equation}
  F^a = (d + V^\al T_\al)^a{}_b\+ X^b\ ,\quad G^\al = dV^\al + \half\+ f_{\be\ga}{}^\al\+ V^\be\! \wedge V^\ga\ .
 \end{equation}
Here, $T_\al{}^a{}_b$ and $f_{\al\be}{}^\ga$ are real representation matrices and structure constants of an arbitrary Lie algebra, satisfying $[\,T_\al\, ,\+T_\be\,]=f_{\al\be}{}^\ga\+T_\ga$, while $g_{ab}=g_{ba}$ and $b_{ab}=(-)^{p+1}b_{ba}$ are arbitrary matrices (the latter nonvanishing only for $2(p+1)=D$). For $D>2$ the action is invariant under two kinds of gauge transformations (neither of which acts on the $V^\al$), but since we are only interested in the case $D=2$ with $p=0$ where the $\t_\al$ and $X^a$ are all scalars, we shall not display them here (see for instance eqs.\ (5) and (6) in \cite{T}). With the gauge symmetries absent, the matrices $g_{ab}$ and $b_{ab}$ may depend on the fields $X^a$.

The $\t_\al$ act as Lagrange multipliers for the flatness conditions $G^\al=0$. Plugging the local solution
 \begin{equation}
  V^\al T_\al = \e^{-\phi\+\cdot T}\# d\+ \e^{\+\phi\+\cdot T} =
  d\phi^\al E_\al{}^\be(\phi)\, T_\be
 \end{equation}
in terms of scalars $\phi^\al$ into the action yields a nonlinear sigma model for $\phi^\al$ and $X^a$. For what follows it is important to note that the local field redefinition
 \begin{equation} \label{XtoY}
  X^a = \big( \e^{-\phi\+\cdot T} \big){}^a{}_b\, Y^b
 \end{equation}
allows to write the field strengths $F^a$ as
 \begin{equation}
  F^a = \big( \e^{-\phi\+\cdot T} \big){}^a{}_b\, dY^b\ ,
 \end{equation}
such that we arrive at the action
 \begin{equation} \label{Ldual}
  \tilde{L}_\si = \frc{1}{2}\, \big( h_{\al\be} *\! d\phi^\al\! \wedge d\phi^\be + \hat{g}_{ab} *\! dY^a\! \wedge dY^b + \hat{b}_{ab}\, dY^a\! \wedge dY^b \big)
 \end{equation}
with the metric $h_{\al\be}(\phi)=E_\al{}^\ga(\phi)\+ E_\be{}^\de (\phi)\,\de_{\ga\de}$ and
 \begin{equation}
  \hat{g}_{ab}(\phi,Y) = \frc{\p X^c}{\p Y^a}\, \frc{\p X^d}{\p Y^b}\, g_{cd}(X)\ ,\quad \hat{b}_{ab}(\phi,Y) = \frc{\p X^c}{\p Y^a}\, \frc{\p X^d}{\p Y^b}\, b_{cd}(X)\ .
 \end{equation}
Now suppose that \eqref{XtoY} is a symmetry of $g_{ab}$ and $b_{ab}$, in particular that the vectors
 \begin{equation} \label{V}
  \V_\al = T_\al{}^a{}_b\+ X^b\+ \p_a
 \end{equation}
satisfy the Killing equation
 \begin{equation} \label{iso}
  \cL_{\V_\al}\+ g_{ab} = T_\al{}^c{}_d\+ X^d\+ \p_c\+ g_{ab} + g_{ac}\+ T_\al{}^c{}_b + g_{bc}\+ T_\al{}^c{}_a = 0\ .
 \end{equation}
Then $\hat{g}_{ab}=g_{ab}(Y)$ and $\hat{b}_{ab}=b_{ab}(Y)$ are independent of $\phi^\al$ and the latter decouple completely from the $X^a$ in the action, as we first observed in \cite{T} (for arbitrary form degree $p$ and dimension $D$).

Alternatively, we may eliminate the auxiliary 1-forms $V^\al$ from the action by means of their algebraic equations of motion, which for $D=2$ read
 \begin{equation}
  K_{\al\be}\+ V^\be = *\+ d\t_\al - (dX^a g_{ab} + *\+ dX^a b_{ab})\, \V^b_\al \equiv *H_\al
 \end{equation}
with the operator
 \begin{equation} \label{K}
  K_{\al\be} = \de_{\al\be} + g_{ab}\+ \V^a_\al\+ \V^b_\be - (f_{\al\be} {}^\ga \t_\ga + b_{ab}\+ \V^a_\al\+ \V^b_\be)\# *\, .
 \end{equation}
Denoting its inverse by $K^{\al\be}$ and substituting the solution for $V^\al$ into the action \eqref{HKmodel}, we arrive at a dual sigma model in terms of $\t_\al$ and $X^a$,
 \begin{equation} \label{genSol}
  L_\si = \frc{1}{2}\, \big(\! *\! H_\al \wedge K^{\al\be} H_\be + g_{ab}\# *\! dX^a\! \wedge dX^b + b_{ab}\, dX^a\! \wedge dX^b \big)\ ,
 \end{equation}
the leading terms of which are given by
 \begin{align*}
  L_\si = \frc{1}{2}\, \big( & \de^{\al\be}\#\# *\! d\t_\al \wedge d\t_\be + g_{ab}\# *\! dX^a\! \wedge dX^b - 2\+ \de^{\al\be} b_{ab}\+ \V^b_\be\+ *\! dX^a\! \wedge d\t_\al \notag \\*
  & + b_{ab}\, dX^a\! \wedge dX^b - f^{\al\be\ga} \t_\ga\, d\t_\al \wedge d\t_\be - 2\+ \de^{\al\be} g_{ab}\+ \V^b_\be\, dX^a\! \wedge d\t_\al + \dots \big)\ .
 \end{align*}
In this way, we have rederived the non-abelian Buscher rules of \cite{GR2} (applied to our models), apart from the dilaton transformation, which is given by the logarithm of the determinant of $K_{\al\be}$ that arises as a Jacobian upon integrating out the $V^\al$ in a path integral \cite{B2,GR1}. It should be noted, however, that T-dual field configurations generically correspond to the same world-sheet theory only in the abelian case \cite{GR2}.

As it is cumbersome to explicitly invert\footnote{Writing $K=S-A*$, the inverse is given by $K^{-1}=(S-A\+S^{-1}\!A)^{-1}(S+A*)\+S^{-1}$.} $K_{\al\be}$ to all orders in the general case, we shall consider commuting matrices $T_\al$ and $b_{ab}=0$ from now on. Then $K_{\al\be}$ is reduced to an ordinary symmetric matrix and the complete action reads
 \begin{align}
  L_\si = \frc{1}{2}\, \big( & G^{\al\be}\# *\!d\t_\al \wedge d\t_\be + (g_{ab} - g_{ac}\+ g_{bd}\+ G^{\al\be} \V^c_\al\+ \V^d_\be) *\! dX^a\! \wedge dX^b \notag \\*
  & - 2\+ g_{ab}\+ G^{\al\be} \V^b_\be\+ dX^a\! \wedge d\t_\al\+ \big)
 \end{align}
with the inverse of $K_{\al\be}$ relabeled as
 \begin{equation} \label{G}
  G^{\al\be} = \big( \de_{\al\be} + g_{ab}\+ \V^a_\al\+ \V^b_\be \big)^{-1}\ .
 \end{equation}

It now follows from the conclusion in section \ref{sec:Tduality} that if $g_{ab}(X^c)$ is an arbitrary $(n-m)$-dimensional Ricci-flat metric with $m$ commuting Killing vectors \eqref{V}, then the following $n$-dimensional field configurations are solutions to the equations of motion \eqref{betaG}--\eqref{betaP}:
 \begin{gather}
  ds^2 = G_{IJ}\, dX^I dX^J = \big( g_{ab} - g_{ac}\+ g_{bd}\+ G^{\al\be} \V^c_\al\+ \V^d_\be \big)\, dX^a dX^b + G^{\al\be} d\t_\al\+ d\t_\be \notag \\[2pt]
  B = g_{ab}\+ G^{\al\be} \V^b_\be\, dX^a\! \wedge d\t_\al\ ,\quad \Phi = \half \ln \det(G^{\al\be})\ , \label{bkg}
 \end{gather}
where we assume that the scalar products of the $\V_\al$  with respect to $g_{ab}$ are compatible with the condition $\det(G^{\al\be})>0$. Regarding the matrices $T_\al$ as coupling constants, these solutions are smooth deformations of the metric $\diag(g_{ab},\de^{\al\be})$ with vanishing matter.

Note that none of the background fields depend on the extra coordinates $\t_\al$. Moreover, the isometries \eqref{iso} of $g_{ab}$ extend to symmetries of the fields in \eqref{bkg}. This follows immediately from the fact that they depend on $X^a$ only through $g_{ab}$ and the components of $\V_\al$ themselves:
 \begin{equation} \label{rotinv}
  \cL_{\V_\al} G_{IJ} = \cL_{\V_\al} B_{IJ} = \cL_{\V_\al} \Phi = 0\ .
 \end{equation}
$G_{IJ}$ is positive-definite if $g_{ab}$ is. In particular, the determinants are related by
 \begin{equation}
  \det(G_{IJ}) = \e^{\+4\Phi} \det(g_{ab})\ ,
 \end{equation}
which implies that the string frame volume element $d^{\+n}\!X\sqrt{G\,} \+\e^{-2\Phi}$ is unchanged by the deformation that leads to the field configurations \eqref{bkg}.

As the examples below demonstrate, these solutions are nontrivial even in the simplest cases when $g_{ab}$ is flat. In general, the metric $G_{IJ}$ has nonvanishing curvature. Moreover, for nonzero $T_\al$ the tensor cannot be pure gauge: expanding $G^{\al\be}$ in powers of $X^a$, the leading terms of the field strength of $B_{IJ}$ read
 \begin{equation*}
  H = dB = - \de^{\al\be} (T_{\be\+ab} - \Gamma_{\!cab}\+ \V^c_\be + \dots)\, dX^a\! \wedge dX^b\! \wedge d\t_\al\ ,
 \end{equation*}
where $\Gamma_{\!cab}$ are the Christoffel symbols of $g_{ab}$. Thus, $H\neq 0$ even for constant $g_{ab}$.

That the fields in $\eqref{bkg}$ solve the equations of motion is a consequence of the duality of sigma models outlined above and the isometry condition \eqref{iso}: As we read off from \eqref{Ldual}, the dualization of $\t_\al$ into $\phi^\al$ followed by the coordinate transformation \eqref{XtoY} yields the block-diagonal metric $\tilde{G}_{IJ}=\diag(g_{ab},h_{\al\be})$ and vanishing tensor field. The dilaton vanishes as well, as follows from the fact that dualization produces the inverse determinant. Now, since the $T_\al$ commute we have $h_{\al\be}=\de_{\al\be}$ and $\tilde{G}_{IJ}$ is Ricci-flat thanks to the Ricci-flatness of $g_{ab}$. The equations of motion \eqref{betaG}--\eqref{betaP} then are obviously satisfied and thus the original backgrounds \eqref{bkg} are solutions as well.

As mentioned in the introduction, the solutions for $m=1$ have been found before in \cite{AG} and \cite{ABDR} by means of a null Melvin twist, which essentially consists of two boosts and two T-dualities with a coordinate shift in between, followed by a double-scaling limit of the transformation parameters. Our derivation appears to be simpler in that no more than one T-duality and coordinate transformation is needed. The cases $m>1$ are not covered by these papers.
\medskip

An alternative and sometimes more convenient formulation of the solutions \eqref{bkg} is obtained by passing from $X^a$ to adapted coordinates $(\rho^\al,u^i)$, $i=1,\dots,n-2m$, such that $\V_\al=\p_{\#\rho^\al}\equiv\p_\al$, which is possible since the $\V_\al$ commute among each other and with $\p_{\t_\al}$. The transformation is of the form
 \begin{equation} \label{coordtrans}
  X^a(\rho, u) = \big( \e^{\,\rho\+\cdot T} \big){}^a{}_b\, f^b(u)\ ,
 \end{equation}
with functions $f^a(u)$ chosen such that the change of coordinates is nonsingular. We then find for the metric
 \begin{align}
  ds^2 & = (g_{\al\be} - g_{\al\ga}\+ G^{\ga\de} g_{\de\be})\, d\rho^\al d\rho^\be + 2 (g_{\al i} - g_{\al\be}\+ G^{\be\ga} g_{\ga i})\, d\rho^\al du^i \notag \\*
  & \quad\, + (g_{ij} - g_{i\al}\+ G^{\al\be} g_{\be j})\, du^i du^j + G^{\al\be} d\t_\al\+ d\t_\be\ , \label{ds_rho}
 \end{align}
for the tensor
 \begin{equation} \label{B_rho}
  B = g_{\al\ga}\+ G^{\ga\be} d\rho^\al\! \wedge d\t_\be + g_{i\al}\+ G^{\al\be} du^i\# \wedge d\t_\be\ ,
 \end{equation}
and for the dilaton
 \begin{equation}
  \e^{-2\Phi} = \det(\de_{\al\be} + g_{\al\be}) = 1 + \text{tr}\+ (g_{\al\be}) + \ldots + \det(g_{\al\be})\ ,
 \end{equation}
where in the last equation we have displayed all terms relevant for the cases $m\leq 2$. Here, all coefficients are independent of both $\t_\al$ and $\rho^\al$, the latter invariances being a consequence of the isometries \eqref{iso} in coordinates $(\rho^\al,u^i)$: $\p_\al\+ g_{ab}=0$. The tensor field strength is easily computed:
 \begin{equation}
  H = \p_i\+ (g_{\al\ga}\+ G^{\ga\be})\, du^i\# \wedge d\rho^\al\! \wedge d\t_\be + \p_i (g_{j\al}\+ G^{\al\be})\, du^i\# \wedge du^j\# \wedge d\t_\be\ .
 \end{equation}

In this formulation, the duality to solutions with vanishing matter fields and Ricci-flat metric can be seen in the following way: We may use the independence of $\rho^\al$ to dualize them into scalars $\si_\al$; from eqs.\ \eqref{Buscher} and \eqref{G} we obtain
 \begin{gather*}
  d\tilde{s}{\+}^2 = g^{\al\be} d\si_\al\+ d\si_\be + (g_{ij} - g_{i\al}\+ g^{\al\be} g_{\be j})\, du^i du^j + \de^{\al\be} (d\t_\al - d\si_\al)\+ (d\t_\be - d\si_\be) \notag \\[2pt]
  \tilde{B} =  g_{i\al}\+ g^{\al\be} du^i\# \wedge d\si_\be\ ,\quad \e^{-2\tilde{\Phi}} = \det(g_{\al\be})\ ,
 \end{gather*}
where $g^{\al\be}$ denotes the inverse of the matrix $g_{\al\be}$, \emph{not} the $\al\be$ components of $g^{ab}$. At this intermediate step it appears that we have to impose the stronger condition $\det(g_{\al\be})>0$. However, the solutions \eqref{ds_rho}, \eqref{B_rho} are perfectly admissible for $\det(g_{\al\be})\leq 0$ as long as $\det(\de_{\al\be}+g_{\al\be})>0$. Now, by introducing shifted coordinates $\phi_\al=\t_\al-\si_\al$, the latter decouple from the other scalars. If we then dualize $\si_\al$ back into $\rho^\al$, it is obvious that we arrive at
 \begin{equation*}
  ds^2 = g_{\al\be}\, d\rho^\al d\rho^\be + 2 g_{i\al}\, du^i d\rho^\al + g_{ij}\, du^i du^j + \de^{\al\be} d\phi_\al\+ d\phi_\be
 \end{equation*}
with $B=\Phi=0$, which is the Ricci-flat metric $\diag(g_{ab}, \de^{\al\be})$ expressed in terms of coordinates $(\rho^\al,u^i, \phi_\al)$.

The actions we have performed here essentially amount to a multi-dimensional version of the so-called TsT transformation \cite{LM,F,I}, which also consists of a coordinate shift sandwiched between two T-duality transformations.

\section{Examples} 
\label{sec:Ex}

\textit{5.1\quad Three-dimensional Euclidean Space} 
\medskip

Let us first consider the simplest case of two coordinates $X^a=(x,y)$ and
 \begin{equation}
  g_{ab} = \delta_{ab}\ ,\quad T_{ab} = -\kappa\+ \eps_{ab}
 \end{equation}
with $\eps_{12}=1$ and constant $\kappa\in\fR$, which obviously satisfy condition \eqref{iso}. The three-dimensional metric $G_{IJ}$ in string frame then takes the form
 \begin{equation}
  ds^2 = \frc{1}{1 + \kappa^2 r^2}\, \big( dx^2 + dy^2 + \kappa^2 (x\+ dx + y\+ dy)^2 + d\t^2 \big)\ ,
 \end{equation}
where $r^2=x^2+y^2$. The rotational invariance \eqref{rotinv} becomes manifest in polar coordinates $X^I=(r,\vphi,\t)$, in which $\V=\kappa\+ (x\p_y- y\p_x)=\kappa\+\p_\vphi$.\footnote{This is a coordinate transformation \eqref{coordtrans} with $\rho=\vphi/\kappa$, $u=r$ and $f^a=(r,0)$.} The metric turns into
 \begin{equation}
  ds^2 = dr^2 + \frc{1}{1 + \kappa^2 r^2}\, \big( r^2 d\vphi^2 + d\t^2 \big)\ .
 \end{equation}
In these coordinates, the corresponding Ricci tensor is diagonal,
 \begin{equation} \label{Ric_ex1}
  R_{IJ} = \frc{2 \kappa^2}{(1 + \kappa^2 r^2)^3}\ \diag \big[ (2 - \kappa^2 r^2) (1 + \kappa^2 r^2)\, ,\, 2 r^2\, ,\, 1 - \kappa^2 r^2\+ \big]\ ,
 \end{equation}
from which we derive the curvature scalar $R=2\kappa^2\+ (5-2\kappa^2 r^2)/(1+\kappa^2 r^2)^2$. The 2-form $B$-field is given by
 \begin{equation}
  B = \frc{\kappa\+ r^2}{1 + \kappa^2 r^2}\, d\vphi \wedge d\t\ .
 \end{equation}
In three dimensions, its field strength is proportional to the volume form,
 \begin{equation}
  H = \frc{2\kappa\+ r}{(1 + \kappa^2 r^2)^2}\, d^{\+3}\#\#X\ ,
 \end{equation}
which implies that
 \begin{equation} \label{HH_ex1}
  H_{IKL} H_J{}^{KL} = \frc{8 \kappa^2}{(1 + \kappa^2 r^2)^2}\, G_{IJ} \ .
 \end{equation}
The dilaton
 \begin{equation}
  \Phi = - \frc{1}{2} \ln(1 + \kappa^2 r^2)
 \end{equation}
has second derivatives
 \begin{equation} \label{ddphi_ex1}
  \nabla_{\!I} \p_J \Phi = \frc{\kappa^2}{(1 + \kappa^2 r^2)^3}\ \diag \big[ \kappa^4 r^4 - 1\, ,\, -r^2\, ,\, \kappa^2 r^2\+ \big]\ .
 \end{equation}
Adding \eqref{Ric_ex1}, \eqref{ddphi_ex1} and \eqref{HH_ex1} with the appropriate prefactors, we find that the anomaly coefficients \eqref{betaG} of $G_{IJ}$ indeed vanish. Likewise, the other two equations of motion \eqref{betaB} and \eqref{betaP} are easily verified.

If $\t$ parametrizes a compact dimension (with period $2\pi$), the action \eqref{S} turns out to be finite.\footnote{In both frames the curvature and dilaton terms diverge separately, but their sum is finite.} Without the boundary term, it is proportional to the flux of the 3-form field strength and negative as anticipated:
 \begin{equation}
  S_3 - S^b_3 = -\! \int\! d^{\+3}\#\#X\, \frc{8 \kappa^2 r}{(1 + \kappa^2 r^2)^2} = - 4 \kappa \int\! H = - 4\+ (2\pi)^2
 \end{equation}
for $\kappa\neq 0$. The boundary term \eqref{Sb}, however, is nonvanishing,
 \begin{equation}
  S^b_3 = 4 \int\! d^{\+3}\#\#X\, \p_I \big( \sqrt{G\,}\+ G^{IJ} \p_J\, \e^{-2\Phi} \big) = 4\+ (2\pi)^2 \int\limits_0^\infty\! dr\ \p_r \Big( \frc{2 \kappa^2 r^2}{1 + \kappa^2 r^2} \Big) = 8\+ (2\pi)^2\ ,
 \end{equation}
and yields a total action with positive value $S_3=4\+(2\pi)^2$.
\bigskip

\textit{5.2\quad Three-dimensional Lorentzian Spacetime} 
\medskip

Instead of a Euclidean metric $g_{ab}$, we can also consider the indefinite metric $g_{ab}=\eta_{ab}=\diag(-1,1)$. The matrix $T_{ab}$ remains the same as in the previous example, but ${T^a}_b=-\kappa\+ \eta^{ac} \eps_{cb}=\kappa\+{\tau_1{\#}^a}_b$ is different\footnote{We denote with $\tau_i$ the Pauli matrices, while $\si_i$ denote the left-invariant 1-forms of SU(2) given below in \eqref{si}.} now, resulting in $\V=\kappa\+(x\p_y+y\p_x)$. The dilaton turns into $\e^{-2\Phi}=1+\kappa^2(x^2-y^2)$. Positivity of this expression restricts the domain of $x,y$ to the region between the two hyperbolas $y^2=\kappa^{-2}+x^2$. We could introduce new coordinates via \eqref{coordtrans}, which would express $x$ and $y$ in terms of linear combinations of $\sinh\rho$ and $\cosh\rho$, such that $\V=\kappa\+ \p_\rho$. Instead, the spacetime geometry becomes more transparent if we use Kruskal-like coordinates $T$ and $X$ defined through
 \begin{equation}
  x = (1+r)\, \e^{-r}\, T\ ,\quad y = (1+r)\, \e^{-r}\+ X\ .
 \end{equation}
Here, the function $r(T,X)$ is determined implicitly by
 \begin{equation} \label{radius}
  \kappa^2\+ \big( T^2 - X^2 \big) = \frc{r-1}{r+1}\, \e^{\+2r}
 \end{equation}
and is restricted by the condition $r>0$, which implies $X^2< \kappa^{-2}+T^2$. It is plotted in figure~1. The string frame metric $G_{IJ}$ then reads
\begin{align}
  ds^2 & = \frc{1}{1 + \kappa^2 (x^2 - y^2)}\, \big( dy^2 - dx^2 - \kappa^2 (x\+ dx - y\+ dy)^2 + d\t^2 \big) \notag \\*[4pt]
  & = \Big( \frc{1 + r}{r} \Big)^{\!2}\, \e^{-2r} \big( dX^2 - dT^2 \big) + \frc{1}{r^2}\, d\t^2\ , \label{g_ex2}
 \end{align}
and the tensor and dilaton are given by
 \begin{equation}
  B = \kappa\+ \Big( \frc{1 + r}{r} \Big)^{\!2}\, \e^{-2r} \big( X\+ dT - T\+ dX \big)\# \wedge d\t\ ,\quad \e^{-2\Phi} = r^2\ .
 \end{equation}

\begin{figure}
\hspace{1em}
\begin{minipage}{7cm}
\begin{center}
\includegraphics[viewport=214 562 384 653]{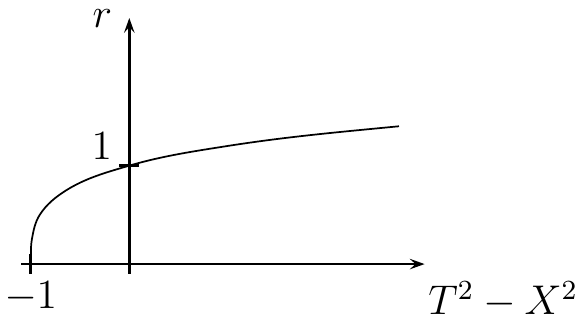} \\[1ex]
{\small Fig.~1: (Inverse) radius \eqref{radius} in (string) Einstein frame}
\end{center}
\end{minipage}
\hspace{1em}
\begin{minipage}{7cm}
\begin{center}
\includegraphics[viewport=236 538 359 677]{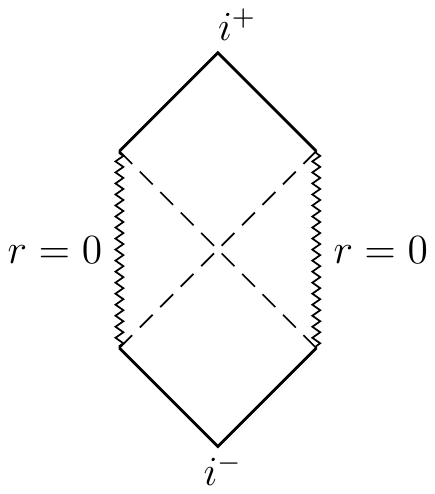} \\[.5ex]
{\small Fig.~2: Penrose diagram of spacetime \eqref{g_ex2}}
\end{center}
\end{minipage}
\end{figure}

Taking $\t$ periodic, the two-dimensional space is cylindrical with varying radius $r^{-1}=\e^{\+\Phi}$ in string frame and $r$ in Einstein frame. In the latter the metric reads $\hat{G}_{IJ}= r^4\+G_{IJ}$. At the two boundaries where $r=0$, the matter fields blow up. So does the curvature scalar, which is given by $R=2\kappa^2(2r^2-7)/r^4$ in string frame and by $\hat{R}=2\kappa^2(5-2r^2)/r^8$ in Einstein frame. The Penrose diagram shown in figure~2 for $\kappa\neq 0$ (with the $\t$ direction suppressed) shares some similarities with that of the interior region of the Reissner-Nordstr\"om black hole, in that there are two \emph{timelike} singularities. The spacetime therefore contains Cauchy horizons.
\bigskip

\textit{5.3\quad Five-dimensional Lorentzian Spacetime} 
\medskip

As a more advanced example, we consider for the Ricci-flat metric $g_{ab}$ the four-dimensional Schwarzschild black hole
 \begin{equation}
  g = - V dt^2 + V^{-1} dr^2 + r^2 \big( d\vt^2 + \sin^2\!\vt\, d\vphi^2 \big) \quad \text{with}\quad V\#(r) = 1 - \frc{2M}{r}\ .
 \end{equation}
There are several choices for an isometry $\V$. Let us pick the time translation $\V=\kappa\+\p_t$ with $\kappa\in\fR$. Positivity of the exponentiated dilaton
 \begin{equation}
  \e^{-2\Phi} = 1 + \kappa^2 g_{tt} = 1 - \kappa^2 V
 \end{equation}
imposes an upper bound on $r$ unless we take $\kappa^2\leq 1$, which we shall assume from now on. Using the coordinates $\rho=t/\kappa$ and $u^i=(r,\vt,\vphi)$, we obtain from \eqref{ds_rho} the five-dimensional string frame metric
 \begin{equation}
  ds^2 = - \frc{V}{1 - \kappa^2 V}\, dt^2 + V^{-1} dr^2 + r^2 \big( d\vt^2 + \sin^2\!\vt\, d\vphi^2 \big) + \frc{1}{1 - \kappa^2 V}\, d\t^2\ ,
 \end{equation}
and from \eqref{B_rho} the tensor
 \begin{equation}
  B = - \frc{\kappa V}{1 - \kappa^2 V}\, dt \wedge d\t\ .
 \end{equation}
The value $\kappa^2=1$ is special: The metric ceases to be asymptotically flat, as indicated by the fall-off behavior of the curvature scalar for large $r$, which in Einstein frame goes like $r^{-4/3}$ only instead of $r^{-4}$ if $0<\kappa^2<1$.

In five spacetime dimensions, the 2-form $B$ can be dualized into a 1-form $A$. Its field strength is given by
 \begin{equation}
  F = - \e^{-2\Phi}\! *\!\# H = \frc{2M\# \kappa}{r^2}\, \e^{\+2\Phi} *\!\# \big( dr \wedge dt \wedge d\t \big) = 2M\# \kappa\+ \sin\#\vt\, d\vt \wedge d\vphi\ .
 \end{equation}
Thus, locally $F=dA$ with $A=-2M\#\kappa\cos\#\vt\,d\vphi$. The flux of $F$ through a 2-sphere at constant $t$, $\t$ and $r>2M$ yields a magnetic charge per unit length
 \begin{equation}
  \frc{1}{4\pi} \int_{S^2}\! F = 2M\# \kappa\ .
 \end{equation}

A constant rescaling of $t$ and $\t$ together with a gauge transformation of $B$ and a shift of the dilaton maps this solution to the five-dimensional black string found in \cite{HHS} with $\cosh^2\!\al =(1-\kappa^2)^{-1}$, which also was obtained from the product of the four-dimensional Schwarzschild black hole with a line/circle, in this case by means of a Lorentz boost mixing $t$ and $\t$ followed by a dualization of $\t$.

There is an easy way to derive an electrically charged black string from the magnetic one: Introduce a tensor $\tilde{B}=A\wedge d\t$, where $A$ is the above 1-form, change the sign of the dilaton, $\tilde{\Phi}= -\Phi$, and rescale the metric by $\e^{-2\Phi}$ with an extra factor $\e^{-2\Phi}$ for $d\t^2$:
 \begin{gather}
  d\tilde{s}^2 = -V dt^2 + \big( 1 - \kappa^2 V \big) \big( V^{-1} dr^2 + r^2 (d\vt^2 + \sin^2\!\vt\, d\vphi^2) + d\t^2 \big) \notag \\[2pt]
  \tilde{B} = -2M\# \kappa \cos\vt\, d\vphi \wedge d\t\ ,\quad \e^{-2\tilde{\Phi}} = \frc{1}{1 - \kappa^2 V}\ . \label{elBH}
 \end{gather}
We will see in the next two sections why this configuration solves the equations of motion (in string frame). Like the Schwarzschild geometry and its magnetically charged cousin the spacetime has a singularity at $r=0$, as can be inferred for $\kappa\neq 0$ from the curvature scalar
 \begin{equation}
  \tilde{R} = \frc{2M^2\# \kappa^2}{r^4}\, \frc{5 - 2\kappa^2 V}{(1 - \kappa^2 V)^3}\ .
 \end{equation}
It is hidden behind a horizon of topology $S^2\times S^1$ (for compact $\t$) at $r=2M$. Integrating the 3-form field strength $\tilde{H}= d\tilde{B}$ over $S^2\times S^1$ at fixed $t$ and $r>2M$ yields an electric charge $2M\#\kappa$.
\bigskip

\textit{5.4\quad Six-dimensional Space} 
\medskip

Let us now give an example for the solutions \eqref{bkg} with two additional dimensions para\-me\-trized by $\t_1$, $\t_2$. We choose as the seed solution the flat metric $g_{ab}=\de_{ab}$ on $\fR^4$. Two commuting Killing vectors \eqref{V} are obtained from the antisymmetric $4\times 4$ matrices
 \begin{equation}
  T_1 = -\I\+ \kappa_1\, \tau_3 \otimes \tau_2\ ,\quad T_2 = -\I\+ \kappa_2\, \mathbbm{1}\# \otimes \tau_2
 \end{equation}
with $\kappa_1,\kappa_2\in\fR$, which span the Cartan subalgebra of so(4). We can introduce adapted coordinates $\rho^\al=(\vphi/2\kappa_1, \psi/2\kappa_2)$ and $u^i=(r,\vt)$ as in \eqref{coordtrans},
 \begin{equation}
  X^a = \e^{\,\vphi\+ T_1 / 2\kappa_1 +\+ \psi\+ T_2 / 2\kappa_2} \lp r \cos(\vt/2) \\ 0 \\ r \sin(\vt/2) \\ 0 \rp = r \lp \cos(\vt/2)\, \cos\! \big( \tfrac{\psi + \vphi}{2} \big) \\[2pt] \cos(\vt/2)\, \sin\! \big( \tfrac{\psi + \vphi}{2} \big) \\[2pt] \sin(\vt/2)\, \cos\! \big( \tfrac{\psi - \vphi}{2} \big) \\[2pt] \sin(\vt/2)\, \sin\! \big( \tfrac{\psi - \vphi}{2} \big) \rp
 \end{equation}
with $r^2=\de_{ab}\+X^a X^b$, in terms of which the metric $g_{ab}$ reads
 \begin{equation}
  g = dr^2 + r^2 \big( \si_1^2 + \si_2^2 + \si_3^2 \big) = dr^2 + \frc{r^2}{4}\, \big( d\vt^2 + d\vphi^2 + 2 \cos\vt\, d\vphi\+ d\psi + d\psi^2 \big)\ .
 \end{equation}
Here, $\si_i$ denote the left-invariant 1-forms of SU(2):
 \begin{align}
  2\si_1 & = \cos\psi\, d\vt + \sin\psi \sin\vt\, d\vphi \notag \\
  2\si_2 & = \sin\psi\, d\vt - \cos\psi \sin\vt\, d\vphi \notag \\
  2\si_3 & = d\psi + \cos\vt\, d\vphi\ . \label{si}
 \end{align}
The Killing vectors turn into $\V_1=2\kappa_1\p_\vphi$ and $\V_2= 2\kappa_2\+\p_\psi$. Since their scalar product with respect to $g_{ab}$ is nonzero, the matrix \eqref{K} has off-diagonal elements,
 \begin{equation}
  K_{\al\be} = \lp 1 + \kappa_1^2\, r^2 & \kappa_1 \kappa_2\+ r^2 \cos\vt \\[2pt] \kappa_1 \kappa_2\+ r^2 \cos\vt & 1 + \kappa_2^2\, r^2 \rp ,
 \end{equation}
which leads to the dilaton
 \begin{equation}
  \e^{-2\Phi} = 1 + (\kappa_1^2 + \kappa_2^2)\+ r^2 + \kappa_1^2\+ \kappa_2^2\+ r^4 \sin^2\!\vt\ .
 \end{equation}
The formulas in section \ref{sec:Sol} now give the six-dimensional metric
 \begin{align}
  ds^2 & = dr^2 + \frc{r^2}{4}\, d\vt^2 + \frc{r^2}{4}\, \e^{\+2\Phi} \big[ (1 + \kappa_2^2\+ r^2 \sin^2\!\vt)\+ d\vphi^2 + 2 \cos\vt\, d\vphi\+ d\psi + (1 + \kappa_1^2\+ r^2 \sin^2\!\vt)\+ d\psi^2 \big] \notag \\*[2pt]
  & \quad\, + \e^{\+2\Phi} \big[ (1 + \kappa_2^2\+ r^2)\+ d\t_1^2 - 2 \kappa_1 \kappa_2\+ r^2 \cos\vt\, d\t_1\+ d\t_2 + (1 + \kappa_1^2\+ r^2)\+ d\t_2^2\+ \big] \label{G6d}
 \end{align}
and tensor
 \begin{align}
  B = \frc{r^2}{2}\, \e^{\+2\Phi} \big[ & \kappa_1 \big(\! \cos\vt\, d\psi +  (1 + \kappa_2^2\+ r^2 \sin^2\!\vt)\+ d\vphi \big) \wedge d\t_1 \notag \\*[-2pt]
  & + \kappa_2 \big(\! \cos\vt\, d\vphi + (1 + \kappa_1^2\+ r^2 \sin^2\!\vt)\+ d\psi \big) \wedge d\t_2\+ \big]\ .
 \end{align}

In six dimensions, the 2-form $B$ can be dualized into another 2-form $\tilde{B}$, which provides us with a further solution. Due to the properties of the Hodge operator in Euclidean signature, we have to consider an imaginary field $\tilde{B}$ defined through $d\tilde{B}=\I\+ \e^{-2\Phi}\!*\!H$ if we want to retain the form of the action \eqref{S}.\footnote{
In Euclidean spaces it is $*^2 H=-(-)^n H$ on forms of odd degree, which differs by a sign from the corresponding expression in Lorentzian signature. For a more detailed explanation of the appearance of the imaginary unit see e.g.\ the appendix of \cite{BCPVV}.} The result then is
 \begin{equation}
  \tilde{B} = \frc{\I}{2}\, r^2 \big[ \kappa_2\+ (d\psi + \cos\vt\, d\vphi) \wedge d\t_1 + \kappa_1\+ (d\vphi + \cos\vt\, d\psi) \wedge d\t_2\+ \big]
 \end{equation}
modulo a gauge transformation. The dualization has to be performed in Einstein frame\footnote{Note that $\hat{*}H=*H$ for a Weyl rescaling \eqref{Weyl} in $n=6$ dimensions.} together with an inversion $\tilde{\Phi}=-\Phi$ of the dilaton in order for $\tilde{\Phi}$ to appear in the action in the same way as $\Phi$. The string frame equations of motion \eqref{betaG}--\eqref{betaP} are then solved with the metric \eqref{G6d} multiplied by a factor $\e^{-2\Phi}$, which is the net effect of two Weyl rescalings $\tilde{G}_{IJ}= \e^{\+\tilde{\Phi}}\hat{G}_{IJ}= \e^{\+\tilde{\Phi}-\Phi}G_{IJ}$ from string to Einstein frame and back.
\bigskip

\textit{5.5\quad More Solutions in Five Dimensions} 
\medskip

A way to turn the solution just derived into one with real fields is to switch off one of the deformations by setting, say, $\kappa_1=0$ and to perform a Wick rotation $\t_1\rightarrow\I\+t$, which results in a spacetime with one timelike direction. We find
 \begin{gather}
  d\tilde{s}^2 = \big( 1 + \kappa^2 r^2 \big) \big(\! -dt^2 + dr^2 + r^2 \si_1^2 + r^2 \si_2^2 \big) + r^2 \si_3^2 + d\t_2^2 \notag \\[2pt]
  \tilde{B} = -\kappa\+ r^2\+ \si_3 \wedge dt\ ,\quad \e^{-2\tilde{\Phi}} = \frc{1}{1 + \kappa^2 r^2} \label{dual5dsol}
 \end{gather}
with $\kappa\equiv\kappa_2$. We may now reduce the spacetime to five dimensions by setting $\t_2$ constant, which still solves the equations of motion. $\tilde{B}$ can then be dualized into a 1-form $\tilde{A}$ again, with field strength
 \begin{equation}
  \tilde{F} = - \e^{-2\tilde{\Phi}}\! *\!\# \tilde{H} = - \kappa\, d\+ \Big( \frc{r^2\+ \si_3}{1 + \kappa^2 r^2} \Big)\ .
 \end{equation}
Note that in a suitable gauge $\tilde{B}=\e^{\+2\tilde{\Phi}}\#\tilde{A} \wedge dt$. The magnetic charge obtained by integrating $\tilde{F}$ over a 2-sphere at constant $t$, $\psi$ and $r$,
 \begin{equation}
  \frc{1}{4\pi} \int_{S^2}\! \tilde{F} = \frc{1}{2\kappa}\, \frc{\kappa^2 r^2}{1 + \kappa^2 r^2}\ ,
 \end{equation}
approaches the finite value $1/2\kappa$ in the limit $r\rightarrow \infty$.
\medskip

What we have just found is a new method to generate solutions in five dimensions, which can be applied to other seed metrics as well. Let us summarize the steps before presenting another example: Starting from a four-dimensional Ricci-flat metric with an isometry, first generate a solution in five dimensions with coordinates $X^I=(X^a,\t)$ using the formulas in section~\ref{sec:Sol}. Then uplift to six dimensions by taking the product with a time- or spacelike factor $\fR$ (or $S^1$) parametrized by $\t'$, depending on whether the seed metric has Euclidean or Lorentzian signature, respectively. Now dualize the tensor (which yields a real field since the six-dimensional metric is already Lorentzian, avoiding the need for a Wick rotation), invert the dilaton and rescale the metric by a factor $\e^{-2\Phi}$. The spacetime can then be reduced to five dimensions with coordinates $X^I=(X^a,\t')$ by setting $\t$ constant. The resulting metric, tensor and dilaton will satisfy the string frame equations of motion \eqref{betaG}--\eqref{betaP}. This is equivalent to the procedure that generated the electrically charged black string \eqref{elBH}. There is no way to derive these solutions by any of the methods mentioned so far (ours, null Melvin twist, TsT transformation), since every one of them acts on the dilaton merely by T-duality transformations, whereas here we also include an S-duality transformation $\Phi\rightarrow \tilde{\Phi}=-\Phi$, which results in $\e^{-2\tilde{\Phi}}$ depending nonpolynomially on the deformation parameter (as evidenced by $\kappa$ appearing in the denominator in \eqref{elBH} and \eqref{dual5dsol}).

Let us apply this method to the Euclidean Taub-NUT metric in the form
 \begin{equation}
  g = V\+ \big( dr^2 + r^2 (2\si_1)^2 + r^2 (2\si_2)^2 \big) + \frc{M^2}{V}\, (2\si_3)^2
 \end{equation}
with $V\#(r)=1+M\#/r$ and constant $M>0$, which is Ricci-flat. We use the Killing vector $\V=\kappa\+\p_\psi$ to generate a new solution in five dimensions. Skipping all further steps described above, the final result for the five-dimensional spacetime and matter fields reads
 \begin{gather}
  d\tilde{s}^2 = \big( V + \kappa^2 M^2 \big) \big( dr^2 + r^2 (2\si_1)^2 + r^2 (2\si_2)^2 \big) + \frc{M^2}{V}\, (2\si_3)^2 - \Big( 1 + \frc{\kappa^2 M^2}{V} \Big)\+ dt^2 \notag \\[2pt]
  \tilde{B} = - \frc{2\kappa\+ M^2}{V}\, \si_3 \wedge dt\ ,\quad \e^{-2\tilde{\Phi}} = \Big( 1 + \frc{\kappa^2 M^2}{V} \Big)^{\!-1}\, .
 \end{gather}
The spatial slices of constant $t$ are equipped with a generalized Taub-NUT metric as considered in \cite{IK}, with the Hopf fiber of $S^3$ $\kappa$-undeformed. The curvature scalar, which in string frame is given by
 \begin{align}
  \tilde{R} = - \frc{\kappa^2 M^4}{2\+ r^4\+ V^4}\, \frc{10 + 7 \kappa^2 M^2\# /\+ V}{(1 + \kappa^2 M^2\# /\+ V)^3}\ ,
 \end{align}
is regular everywhere. Its fall-off behavior $\sim r^{-4}$ for large $r$ is the same in both frames. There is again a gauge in which the 1-form $\tilde{A}$ dual to $\tilde{B}$ satisfies the relation $\tilde{B}= \e^{\+2\tilde{\Phi}}\#\tilde{A}\wedge dt$. It gives rise to a magnetic charge
 \begin{equation}
  \frc{1}{4\pi} \int_{S^2}\! \tilde{F} = \frc{\kappa\+ M^2}{1 + \kappa^2 M^2 + M\#/r}
 \end{equation}
that is finite for any $r$ and approaches $\kappa\+M^2\#/(1+\kappa^2 M^2)$ in the limit $r\rightarrow\infty$.

\section{Conclusions} 
\label{sec:Con}

We have presented two methods to generate solutions to dilaton gravity coupled to a 2-form gauge potential in $n$ dimensions that is part of the low-energy effective description of (noncritical) bosonic string theory. The primary method generalizes the one-parameter solutions produced by the null Melvin twist and the TsT transformation (restricted to the Neveu-Schwarz sector considered here) to multi-parameter deformations of the seed solution. Of course, one may perform several consecutive twists, each with its own parameter, but the result will in general differ from ours. Note also that the null Melvin twist requires the seed solution to admit at least three commuting Killing vectors, one more than our method. Moreover, in suitable coordinates our method makes use of only one T-duality, whereas the others each involve two such transformations. This is a curious feature because when embedded into a type II supergravity, one T-duality would map from IIA to IIB or vice versa, whereas after two T-dualities one ends up in the same theory.

The reader may have noticed the absence of four-dimensional examples. The reason for this is that, since in three dimensions the curvature tensor is completely determined by the Ricci tensor, there are no Ricci-flat metrics other than flat ones that could be used as seed solutions for four-dimensional spacetimes --- our method is most suited for higher dimensions.

Five dimensions are particularly interesting, not only because of the relation to Einstein-Maxwell-dilaton gravity and the availability of many possible Ricci-flat seed metrics, but also because in this case we have found a further way to generate solutions. What makes this second method special is that it `dualizes' a 2-form gauge potential into another 2-form, whereas the proper dual in five dimensions is of course a 1-form. Furthermore, since it acts on the dilaton by an S-duality transformation, in a string-theoretic context it can be used to derive solutions with small string coupling constant $g_s= \e^{\+\Phi_\infty}$ if the original solution has large $g_s$ and would receive strong quantum corrections. It would be interesting to generalize this method to supergravity theories that include a Ramond-Ramond sector.
\medskip

\textbf{Acknowledgments} \\ 
I would like to thank Thomas Strobl and Sebastian Uhlmann for helpful discussions.

\raggedright 

\end{document}